\documentclass[prb,twocolumn,citeautoscript,superscriptaddress]{revtex4-2}

\usepackage{times}
\usepackage{graphicx}
\usepackage{epsfig}
\usepackage{dcolumn}
\pdfoutput=1

\usepackage{amsmath}

\usepackage{natbib}

\begin{document}
\title{Multimode Jahn-Teller Effect in Negatively Charged Nitrogen-Vacancy Center in Diamond
}

\author{Jianhua Zhang}  \affiliation{School of Physics and Optoelectronic Engineering, Hainan University, Haikou, 570228, China}
\affiliation{Ames Laboratory-U.S. DOE and Department of Physics and Astronomy, Iowa State University, Ames IA 50011, USA} 
\author{Jun Liu} \affiliation{Ames Laboratory-U.S. DOE and Department of Physics and Astronomy, Iowa State University, Ames IA 50011, USA}
\author{Z. Z. Zhu} \affiliation{Collaborative Innovation Center for Optoelectronic Semiconductors and Efficient Devices, Department of Physics, Xiamen University, Xiamen 361005, China}
\author{K. M. Ho} \affiliation{Ames Laboratory-U.S. DOE and Department of Physics and Astronomy, Iowa State University, Ames IA 50011, USA}
\author{V. V. Dobrovitski} \affiliation{QuTech and Kavli Institute of Nanoscience, TU Delft, P.O. Box 5046, 2600 GA Delft, Netherlands}
\author{C. Z. Wang} \affiliation{Ames Laboratory-U.S. DOE and Department of Physics and Astronomy, Iowa State University, Ames IA 50011, USA}

\begin{abstract}
We present a first-principles study of the multimode Jahn–Teller (JT) effect in the excited $^{3}E$ state of the negatively charged nitrogen–vacancy (NV) center in diamond. Using density functional theory combined with an intrinsic distortion path (IDP) analysis, we resolve the full activation pathways of the JT distortion and quantitatively decompose the distortion into contributions from individual vibrational modes. We find that multiple vibrational modes participate cooperatively in the JT dynamics, giving rise to a shallow adiabatic potential energy surface with low barriers, consistent with thermally activated pseudorotation. The dominant JT-active modes are found to closely correspond to vibrational features observed in two-dimensional electronic spectroscopy (2DES), in agreement with recent ab initio molecular dynamics simulations. Our results establish a microscopic, mode-resolved picture of vibronic coupling in the excited-state NV center and provide new insight into dephasing, relaxation, and optically driven dynamics relevant to solid-state quantum technologies.

\end{abstract}

\maketitle

The negatively charged nitrogen-vacancy (NV) center in diamond has attracted significant attention owing to its potential applications in a wide range of technologies\cite{Atature18,Awschalom18}, including quantum information\cite{Rondin14,Hanks17}, magnetometry\cite{Barry20,Qiu21}, and photonics\cite{Fotso16,Buterakos17}. It also provides an excellent platform for investigating fundamental problems in quantum mechanics such as the dynamics of quantum spins coupled to their environment\cite{Lillie17,Abobeih18}, as well as exploring quantum control of solid-state spins\cite{Dobrovitski13,Unden18}.

Lattice vibrations play a crucial role in properties of the NV center. Vibrational modes participate in the optical transitions and give rise to visible phonon sidebands in the absorption and photoluminescence spectra\cite{Davies74}. Phonons further influence electronic relaxation\cite{Fuchs10}, spin decoherence\cite{Rand94}, and photon depolarization\cite{Kai-Mei09}. In particular, electron-phonon coupling induces Jahn-Teller (JT) distortion in the $^{3}E$ excited state of the NV center, significantly impacting its dynamics\cite{Kai-Mei09,Abtew11}. The resulting dynamical Jahn–Teller effect (dJTe) is the dominant mechanism underlying optical dephasing\cite{Kai-Mei09} and electronic depolarization dynamics\cite{Ulbricht16} .

Theoretical modeling using multiphonon models\cite{Ma12} and accurate first principles calculations\cite{Alkauskas14, RN245} have been employed to reproduce the vibronic bands and successfully describe the luminescence and adsorption lineshapes of the NV center. Complementary ab initio studies investigated localized or quasilocalized vibrational modes\cite{Gali11,Zhang11}, confirmed the dynamical Jahn–Teller effect in the excited state\cite{Abtew11}, and attributed multiple intersystem crossing rates at cryogenic temperatures to the dJTe\cite{Thiering17}.

Previous ab initio calculations for the dJTe in the excited state of the NV center treated the system as an ideal $E \bigotimes e$ JT problem, considering only a single doubly degenerate $e$ vibrational mode in the distortion process\cite{Abtew11,Thiering17}. However, a diamond crystallite with an embedded defect center, comprising $N$ atoms in total, has $3N-3$ vibrational degrees of freedom, which can be conveniently described within the harmonic approximation by normal coordinates transforming according to the irreducible representations of the point group symmetry. When $N$ is sufficiently large, more than one vibrational mode may actively participate in the JT distortion, giving rise to the so-called multimode Jahn–Teller problem\cite{Bersuker06}. These active modes may differ in character and play distinct roles in the dynamical processes. It is therefore reasonable to expect that the active vibrational modes directly influence the system dynamics and, consequently, its electronic, spin, and optical properties. A detailed understanding of the multimode JT effect, particularly the dynamical path of JT distortion, is thus essential and warrants further investigation. Recently, based on a finite-supercell vibrational-mode analysis, we employed an intrinsic distortion path (IDP) method to study the multimode JT effect in the ground state of the neutral NV center\cite{Zhang18}. The central idea of the IDP method is to represent the JT-active distortion as a linear combination of normal modes computed at the low-symmetry energy minimum. This approach enables the identification of vibrational modes that make dominant contribution to the distortion. In the present work, we combine density functional theory(DFT) calculations with the IDP method to investigate multimode Jahn–Teller distortions in the NV center in diamond. We find that several vibrational modes contribute significantly to the distortion, and these modes are strongly correlated with the vibrational modes coupled to the NV center observed in two-dimensional electronic spectroscopy (2DES) measurements\cite{Huxter13,Carbery2024}.

The NV center exhibits $C_{3v}$ symmetry in its $^{3}A_{2}$ ground state. Its defect-related electronic structure comprises two fully symmetric singlet $A_{1}$ states ($u \bar{u} v \bar{v}$) and one doubly degenerate $E$ state ($e_{x} \bar{e} _{x} e_{y} \bar{e}_y$), with six electrons occupying $u^{2} v^{2} e^{2}$  in the ground state\cite{Manson06}. The $^{3}E$ excited state is achieved by promoting an electron from a singlet $\bar{v}$(or $v$) to the doublet $\bar{e}_{x}$  or $\bar{e}_{y}$ state, resulting in a $u^{2}v^{1} e^{3}$ configuration (Fig. 1). In this configuration, partial occupation of the degenerate E state gives rise to a Jahn–Teller instability. Consequently, the system undergoes a symmetry-lowering distortion to a more stable nondegenerate geometry with ($C_{1h}$) symmetry.

\begin{table}[htbp]

	\caption{\label{tab:hyperf}  
		Ab initio parameters of APES. $R_{JT}^{M}$ ($R_{JT}^{S}$) denotes the magnitude  of the JT distortion vector from a minimum (saddle point) to the high symmetry ($C_{3v}$) configuration. $E_{JT}^{M} (= E_{HS}^{M}-E_{LS}^{M}$) and  $E_{JT}^{S} (=E_{HS}^{S}-E_{LS}^{S}$)  are the Jahn–Teller stabilization energies of a minimum and a saddle point relative to the high-symmetry structure. $\Delta$ represents the energy barrier between $G_{M}$ and $G_{S}$. 
	}
	\begin{ruledtabular}
	 \begin{tabular}{cccccc}
		\ &  $R_{JT}^{M}(\AA)$   & $E_{JT}^{M}(meV)$  & $R_{JT}^{S}(\AA)$   & $E_{JT}^{S} (meV)$  & $\Delta (meV)$ \\
		\hline	
		HSE06  &  0.0699      &     41.0   &  0.0753 &	38.5  &	9.9	 \\
		
	\end{tabular}
\end{ruledtabular}
\end{table}

Density functional theory calculations were performed using the projector augmented-wave (PAW) method with spin polarization,as implemented in VASP\cite{Kresse96}.The screened hybrid functional HSE06 was employed for geometry optimizations and adiabatic potential energy surface (APES) calculations.  A plane-wave kinetic energy cutoff of 400 eV and the $\Gamma$-point sampling were used. A 216-atom ($3 \times 3 \times 3$) cubic supercells with a box length of $3a_{0} =10.632 {\AA}$ were adopted, where $a_{0}$ is the DFT-optimized lattice constant of diamond. The standard HSE06 parameters were used, with a screened Fock exchange fraction $a = 0.25$ and screening parameter $\mu =0.2$\cite{Heyd03}.

The high-symmetry (HS) $C_{3v}$ configuration of the NV center in the excited state is obtained by  geometry optimization with equal fractional occupation of the two degenerate  $\bar{e}$ orbitals, ($\bar{e} _{x})^{0.5} (\bar{e}_{y})^{0.5}$ (Fig. 1, middle). Starting from this structure, total energies $E^{M}_{HS}$ and $E^{S}_{HS}$ were calculated using integer occupations $(\bar{e}_{x})^{1}(\bar{e}_{y})^{0}$  and $(\bar{e}_{x})^{0} (\bar{e} _{y})^{1}$, corresponding to the lower and upper branches of the excited state, respectively. The small energy difference between these two configurations originates from the approximate nature of DFT\cite{Daul97,Zlatar09}. Subsequent geometry optimization were carried out under $C_{1h}$ symmetry constraints, yielding low-symmetry (LS) configurations. The resulting APES consists of three equivalent minima ($G_{M}$) with electronic occupation $(\bar{a}^{''})^{1} (\bar{a} ^{'})^{0}$ and three saddle points ($G_{S}$) with electronic occupation $(\bar{a}^{'})^{1} (\bar{a} ^{''})^{0}$ as shown in Fig. 2(b). The corresponding energies $E^{M}_{JT}$ and $E^{S}_{JT}$ were obtained, and the energies at other points on the APES were determined by varying the ionic coordinates. In the HS geometry ($G_{HS}$) the three carbon atoms adjacent to the vacancy are equivalent and form an equilateral triangle (Fig. 1: middle). In contrast, in the LS configurations, the C–C bonds between two of these atoms is elongated or compressed (Fig. 1: right and left), resulting in two geometrically distinct isosceles triangles.
The calculated JT parameters can be found in Table I. 
\begin{figure}
      \includegraphics[width=8.6cm,viewport=-0 0 868 795]{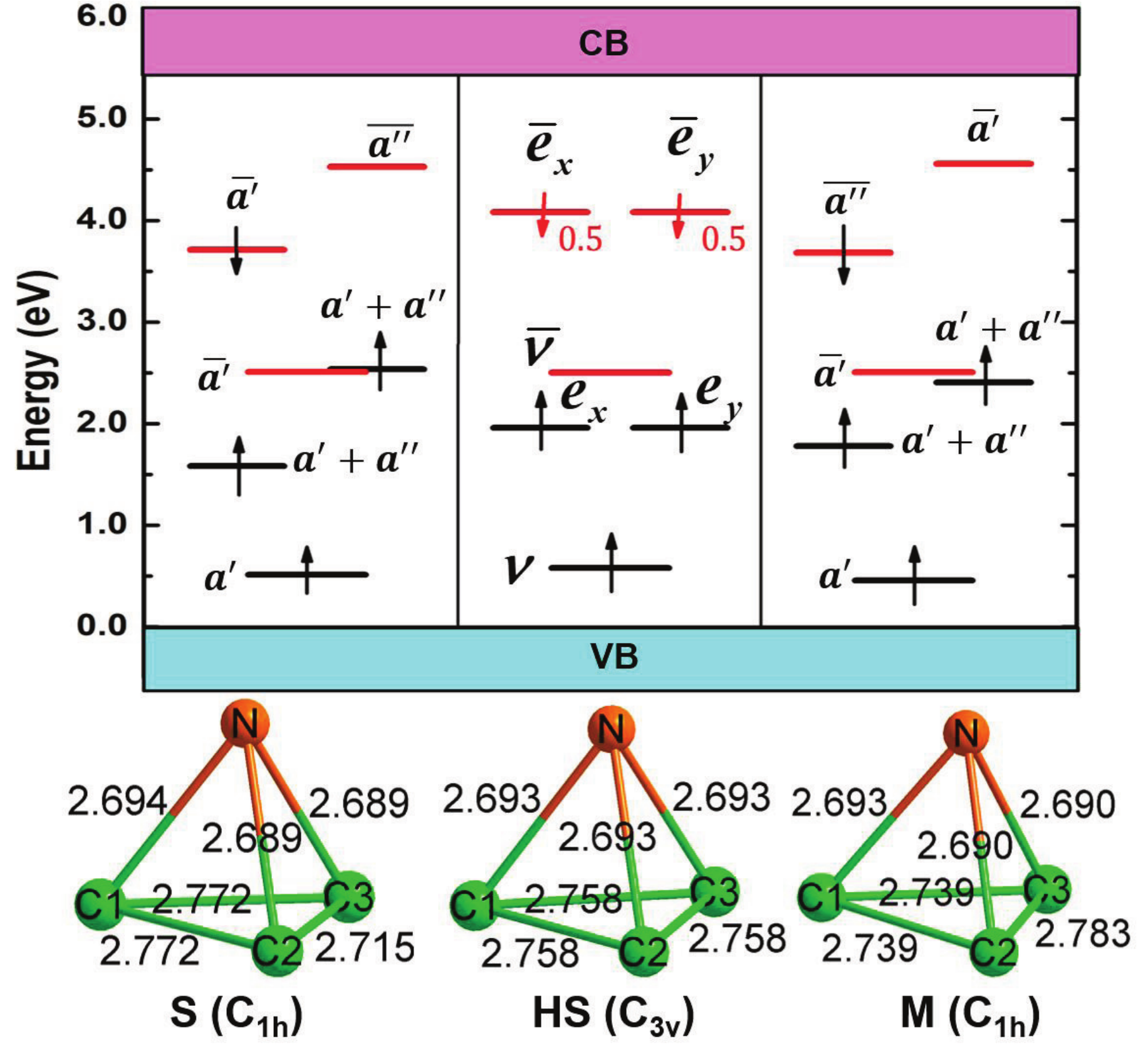}
	\caption{\label{fig1}(Color online) Atomic structure (lower panel) and
		relevant single-electron orbitals (upper panel) of negatively
		charged NV-center in the excited state: the saddle-point configuration (S, left),
		the high symmetry  $C_{3v}$ configuration (HS, middle), and the distorted
		low-symmetry $C_{1h}$ minimum configuration (M, right). As the geometry distorts from HS to M (or S), the degeneracy of electronic configuration $(\bar{e} _{x})^{0.5} (\bar{e}_{y})^{0.5}$ is lifted, yielding the configuration
		 $(\bar{a}^{''})^{1} (\bar{a} ^{'})^{0}$ for M and   $(\bar{a}^{'})^{1} (\bar{a} ^{''})^{0}$ for S. The Jahn-Teller distortion splits the doubly degeneracy of the $e$ states into symmetry-adapted $a^{'}$ and  $a^{''}$ components (denoted as $a^{'} + a^{''}$) in the distorted $C_{1h}$ configurations. In the lower panel, only first neighbor C (green sphere) and N (yellow sphere) atoms surrounding  the vacant are shown. The distances in~\AA\ between the first neighbor atoms
		are also shown. In the upper panel, only defect states between the 	valence band (VB) and conduction band(CB) are shown, with nergies referenced to the valence-band maximum (VBM). The symbols with the bar denote the spin-down states, while unbarred symbols correspond to spin-up components. The states $\nu$ and $\bar\nu$ have $a_1$ symmetry, whereas all other states have $e$ symmetry.}
\end{figure}

To analyze the electron-phonon coupling, the electronic Hamiltonian $H$ of the system is expanded as a Taylor series around $G_{HS}$ with respect to the $3N-3$ normal vibrational coordinates $Q_{HSk}$. The corresponding matrix element can be written as \cite{Zlatar09}
$H_{ij}=E_{0} \delta_{ij}+\sum_{k=1}^{3N-3}\sum_{i,j}^{f}F_{ij}^{k} Q_{HSk}+1/2 \sum_{k=1}^{3N-3}\sum_{i=1}^{f}G_{ii}^{kk} Q_{HSk}^{2} +1/2 \sum_{k,l=1;k \neq l}^{3N-3}\sum_{i,j=1;i\neq j}^{f}G_{ij}^{kl} Q_{HSk}Q_{HSl}+...$,     (1)
where $E_{0}$ is the energy of electronic Hamilton of the $G_{HS}$, $F_{ij}^{k}$, $G_{ii}^{kk}$, $G_{ij}^{kl}$ are the vibronic coupling constants, and the index $f$ indicates the $f$-fold degeneracy of the electronic state.
The JT distortion from the HS geometry towards a LS geometry corresponding to an energy minima can be viewed as a displacement on the adiabatic potential energy surface. Such a distortion generally involves multiple vibrational modes, although it can sometimes be approximated by a single effective mode\cite{Bersuker06}. For the NV center in the excited state with the $C_{3v}$ symmetry, the two-fold degenerate $E$ electronic state couples to vibrational modes according to the  $E \bigotimes e$ scheme,  leading  to the splitting $E \bigotimes e=a1\oplus[a2]\oplus e$.  Only the asymmetric vibrational displacements drive the system away from the HS $C_{3v}$ configuration toward to the LS $C_{1h}$ structures. In general, coupling between degenerate electronic states and degenerate vibrational modes lowers the symmetry of a Jahn–Teller–active system, lifts the electronic degeneracy, and generates two adiabatic potential energy sheets connected by a conical intersection. In the present case, the JT effect splits the doubly degenerate $^{3}E$ state into two energy sheets, with the lower sheet exhibiting a characteristic tricorn-shaped topology (Fig. 2), consistent with the standard $E\bigotimes e$ picture. Our calculations further support the  presence of a dJTe arising from the excited $^3E$ state of the NV center. The adiabatic potential minima are shallow ($E_{JT}^{M}=41 \ meV$), rendering the distorted LS configurations difficult to observe at room temperature\cite{Bersuker06}. The small energy barriers between neighboring minima ($\triangle =9.9\ meV$) enable near-free pseudorotation at ambient temperature, fully consistent with the $E\bigotimes e$ Jahn–Teller framework. 
\begin{figure}
	\includegraphics[width=6.4cm,viewport=20 10 310 490]{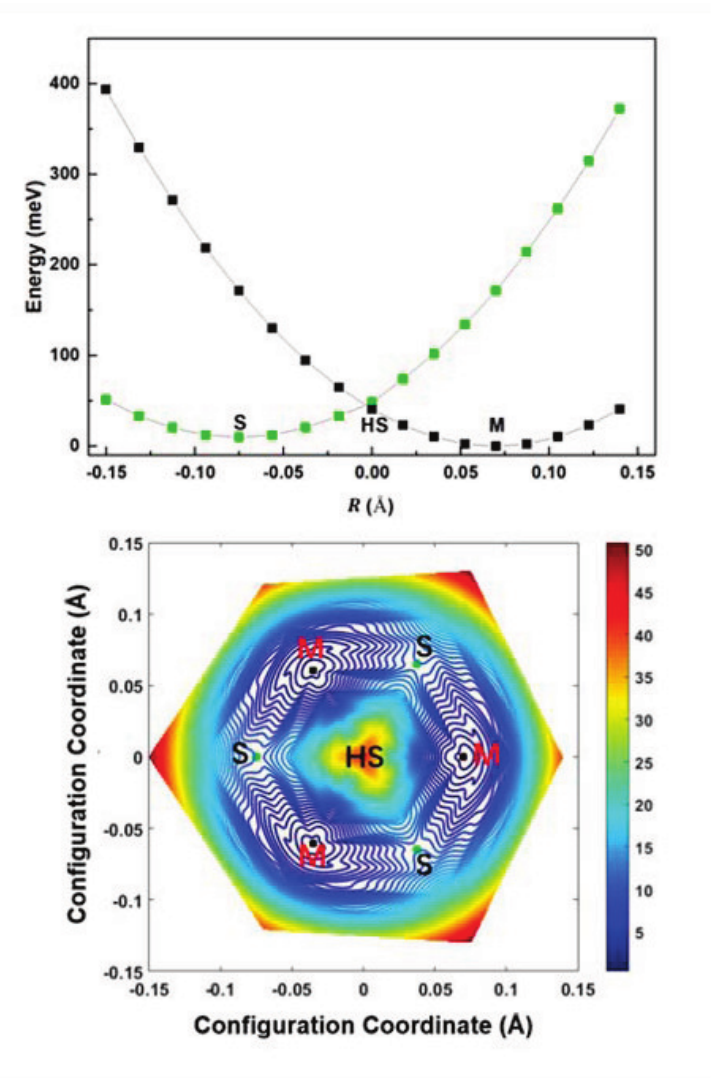}
	\caption{\label{fig2} (Color online) Adiabatic potential energy surface (APES) of the negatively charged NV-center along $R_{JT}^{M}$ (positive) and $R_{JT}^{S}$ (negative) directions in the excited states via DFT calculations. All the points (red points) shown in the figure are ab initio energies calculated from HSE06. The black solid lines are the quadratic fit of the ab initio energies based on Eq. (2). The energy shown in the vertical axis is with respect to the energy minimum point (M).  (b) Contour plot of the lower sheet of the $^3E$ excited states on APES in the surface formed by $\vec{R}_{JT}^{M}$ and $\vec{R}_{JT}^{S}$. The position of minima M (black points) and saddle points S (green points) are also shown. The high symmetry point corresponding to the $G_{HS}$ is at the center. The energy (in unit meV) shown in the figure is with respect to the energy of the $G_{M}$. }
\end{figure}

To further analyze the multimode JT problem of the NV center, we employ the IDP method, using the LS configuration as the reference geometry. The IDP method\cite{Zlatar09,Zlatar10} relies on the fact that all essential information describing the vibronic coupling is encoded in the distorted LS minimum energy geometry $G_{M}$. Accordingly,  $G_{M}$ is chosen as a reference point,  and the distortion can be expressed as a linear combination of all normal modes of the quadratic potential well around $G_{M}$.
Within the harmonic approximation, the energy of a arbitrary point X on the APES in the vicinity of an energy minimum can be represented using a quadratic energy surface\cite{Zlatar09,Zlatar10}:
$E_{X}=\sum_{k=1}^{N_{a1}}E_{k}=1/2\sum_{k=1}^{N_{a1}}w_{Xk}^{2}v_{k}\vec{Q}_{k} \cdot \vec{Q}_{k}$,      (3)
where $\vec{Q}_{k}$ and $\nu_{k}$ are the eigenvectors and frequencies of the vibrational modes in $G_{M}$. $w_{Xk}$ are weighing factors and $N_{a1}$ is the number of vibrational normal modes of $G_{M}$ with nozero weighting factor $w_{Xk}$. The energy at the minimum $G_{M}$ is set to zero, so that the energy at point X represents the vibrational energy relative to the minimum. The vibrational modes are obtained using the finite-displacement method within density functional theory at the PBE level\cite{Alfe09}. Due to their substantially higher computational cost, hybrid functionals are not used for phonon calculations. Previous studies have demonstrated that the relevant vibrational modes are nearly identical at the PBE and HSE levels\cite{Alkauskas14}.

Starting from an arbitrary point X, the system is driven step by step toward the energy minimum along a path determined by the total driving force, which is constructed as a superposition of the forces associated with the individual normal modes\cite{Zlatar09,Zlatar10}:
$\vec{F}_{Xtot}=\sum_{k=1}^{N_a1}\vec{F}_{Xk}=1/2\sum_{k=1}^{N_a1}w_{Xk}v_{k} M^{1/2} \vec{Q}_{k}$,                                      (4)
where $\vec{F}_{Xk}$ are the forces induced by different normal modes which are obtained from the derivative of energy over the Cartesian coordinates of the normal modes, and $M$ is a diagonal $3N \times 3N$ mass matrix with diagonal elements $(m_1, m_1, m_1, m_2, m_2, m_2, \ldots, m_N, m_N, m_N)$, reflecting the three Cartesian degrees of freedom of each atom\cite{Zlatar10}. This path determined by the forces in Eq. (4) is called the intrinsic distortion path (IDP). The distortion vector $\vec{R}_{x}$ (from the point X to the energy minimum point M) can be expressed in terms of the normal modes in $G_{M}$\cite{Zlatar09,Zlatar10}, $\vec{R}_{x}=\sum_{k=1}^{N_a1}w_{Xk}\vec{Q}_{x}$. (5)           In particular if $X$ is exactly on the HS unstable point in the cusp of the lower sheet of APES, the distortion $\vec{R}_{x}$ becomes the JT distortion $\vec{R}_{JT}$, while the corresponding energy is identified as the JT stabilization energy ($E_{JT}$). 

The IDP analysis provides further insight into the vibronic coupling in the NV center. The potential energy profile along the IDP, connecting the cusp to the global minimum (Fig. 3(a)), delineates two different regions. In the first region, the energy decreases rapidly, with more than $40\%$ of the total JT stabilization energy gained within the initial portion of the path. In the second region, the system undergoes a gradual relaxation toward the global minimum. The IDP differs slightly from the direct path (DP), which is defined as a straight path along the JT distortion vector $\vec{R}_{JT}^{M}$ on the APES. The energies along the DP were calculated in the interaction mode framework using equations (3) and (5). Compared to the DP, the IDP is steeper, indicating a faster and stronger stabilization of the system along the physically relevant distortion pathway(Fig. 3(a)). The energies obtained from first-principles calculations along both the IDP and DP are in good agreement with those predicted by the harmonic approximation, except in the immediate vicinity of the HS configuration. Notably, the JT energy obtained from the IDP method ($37.7\ meV$) is slightly lower than the value extracted from direct DFT calculation  ($E_{JT}^{M}=41.0\ meV$). This difference can be partially attributed to anharmonic effects neglected in the harmonic IDP framework\cite{Zlatar10,Ramanantoanina13}. Consistently, the energy barrier $\triangle$ between adjacent minima obtained within the harmonic approximation (approximately $29.8\ meV$) is significantly higher than the corresponding DFT value ($\triangle =9.9\ meV$) (Fig.3(b)). This discrepancy highlights the important role of anharmonicity in softening the adiabatic potential energy surface, which is  captured in fully self-consistent DFT calculations but absent in the harmonic IDP treatment.
\begin{figure}

    \includegraphics[width=8.6cm, viewport=-5 0 555 460]{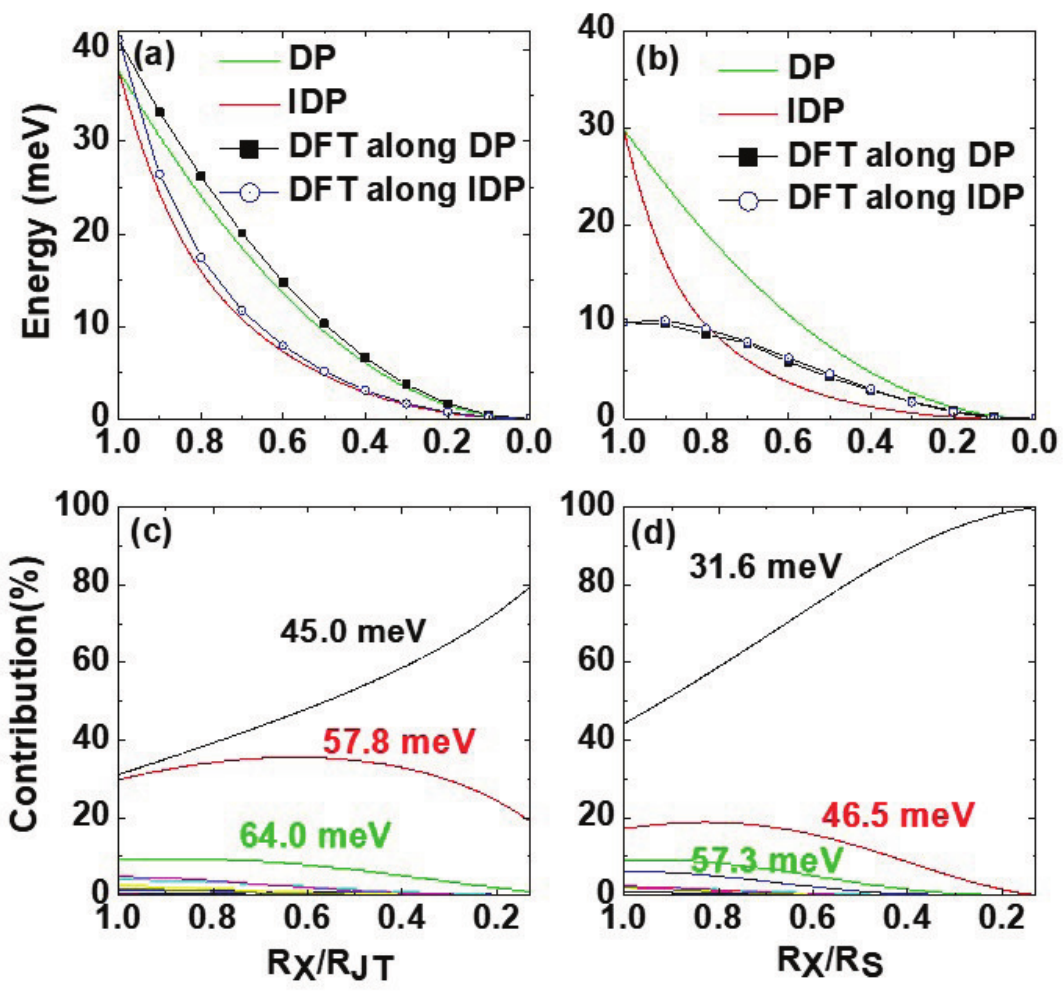}
	\caption{\label{fig3} (Color online) (a, b) Energy evolution from (a) the  $G_{HS}$ ($R_{X}/R_{JT} = 1$) and (b) the $G_{S}$ ($R_{X}/R_{S} = 1$) to the $G_{M}$ ($R_{X}/R_{JT} = 0$) along the direct path (DP) and IDP.  The DP is defined as the straight-line distortion in configuration space along the Jahn-Teller distortion vectors  in (a) $\mathbf{R}_{JT}^M$  and (b) $\mathbf{R}_{JT}^S$ , whereas the IDP corresponds to the force-driven relaxation path determined by the total harmonic forces in Eq.(4). (c, d) Normalized contributions of the dominant vibrational modes to the
		distortion along the IDP from  (c) the $G_{HS}$ ($R_{X}/R_{JT} = 1$) and (d) the $G_{S}$ ($R_{X}/R_{S} = 1$) to the $G_{M}$ ($R_{X}/R_{JT} = 0$).
	}
\end{figure}

\begin{table*}[tbp]
	\caption{\label{tab:hyperf}  Analysis of the multimode JT effect in the excited state of NV center. Modes I (II) denote vibrational modes with large contributions to the JT distortion from the high symmetry geometry (the saddle point) to the energy minimum. $c_{r,m}^{k}$ and $c_{r,s}^{k}$ represent the contributions of the selected normal modes to the distortions $\vec{R}_{JT}^{M}$ and $\vec{R}_{JT}^{S}$, respectively, $c_{e,m}^{k}$ and $c_{e,s}^{k}$ denote their contributions to the energies $E_{M}$ and $E_{S}$. Numbers in parentheses in the second and fifth columns indicate the number of vibration modes with  closely spaced frequencies. Results from 2DES experiment(\cite{Huxter13} / \cite{Carbery2024}) and from AIMD calculation at 10K/77K \cite{Ulbricht16} are included for comparison.}
	\begin{ruledtabular}
		\begin{tabular}{ccccccccc}
			Modes I(meV) &  $c_{r,m}^{k}$ (\%)  & $c_{e,m}^{k}$ (\%)& Modes II (meV) & $c_{r,s}^{k}$ (\%) & $c_{e,s}^{k}$  & 2DES (meV) & AIMD (meV)\\
			\hline
			&        &                    & 31.6   &  43.6(1) &  11.0	& 32.1±2.4 /36.3 &	35.7 / 32.0 \\	
			45.0±0.8 & 31.1(3) &   11.7  &  46.5±1.8  &	17.0(5) &	9.5	& 41.8±3.8 /47.5	& 46.8 / 46.9 \\
			57.8±2.4 & 29.8(5)	 & 15.0  &	57.3±2.0 &	8.6(9) & 6.7	 & 56.7±2.2 &	55.2 / 57.7 \\
			64.0±0.5 & 9.1(4)	& 7.0	  & 63.4±1.9 &	6.3(6)	& 6.3 & & & \\           
			66.1±0.5 &	1.2(2)	& 1.0	 & 66.3±0.3  &	1.9(2)	& 2.1	 & 69.2±2.0	/66.2 & 66.2 / 68.6 \\
			75.3±1.3 & 3.9(5)	& 4.1	& 75.9 ±0.7 & 	2.2(3)	&  3.3	&  / 75.0 & / 76.8   \\
			81.8±0.8 &	4.8(5)	& 5.4	& 81.4±1.1	& 1.8(5) &	3.1 & 79.8±2.6 &   79.9 /   \\
			86.5±0.7 &  2.4(2)  & 2.9   & 86.8±1.5  & 1.9(5)       & 3.7     &  /85.3     &   88.1 / 88.1    \\
			94.9±0.9 &  1.7(4)  &       &     95.3±0.1  &1.0(2) & 2.3 & 91.3±2.2   & 93.6 / 93.3 \\
			 106.5±0.1	& 0.6(2)	& 1.3	&	       &	&	& 105.7	 & 110.2 / 107.2       \\
			         &           &    &     127.9     &0.3(1) & 1.1   &   127.5±2.7	 & 126.7 / 126.8 \\		  
	       147.9       & 0.3(1)	& 	&	  &	 &                  	& 144.6±2.7	& 146.0 / 148.5 \\
	       157.3	& 0.4(1) &	1.9	        & 156.9 &	0.6(1)	& 3.6	 & 156.5	& 154.4 / 156.7 \\ 	
	     164.5  &	1.6(1) &	8.1	 & 164.0±0.9 &	1.5(2)	& 9.9 &	166.0±2.7 &  159.8 / 162.4 \\     
	     Total	& 87.0 & 64.7 &   &	87.9  &	62.5   &  &
	      	
		\end{tabular}
	\end{ruledtabular}
\end{table*}

Within the harmonic approximation, the contributions of individual normal modes to the Jahn–Teller (JT) distortion and their evolution along the intrinsic distortion path (IDP) can be quantitatively analyzed. Specifically, in $G_{M}$ several $a1$ modes ($45.0\ meV$, $57.8\ meV$, and $64.0\ meV$) dominate the lattice distortion from $G_{M}$ to $G_{HS}$,contributing approximately $31.1\%$, $29.8\%$, and $9.1\%$, respectively(Fig. 3c). While for the distortion from $G_{S}$ to $G_{M}$, the dominant modes at $31.6\ meV$, $46.5\ meV$ and $57.3\ meV$, account for $43.6\%$, $17.0\%$, and $8.6\%$ of the total distortion, respectively (Fig. 3d). The relative contributions of these modes vary continuously along the IDP (Fig. 3(c, d)), demonstrating that the importance of each mode in the JT distortion depends on the distance away from $G_{HS}$ (or $G_{S}$). Interestingly, the contributions of the most significant modes ($45.0\ meV$ in Fig. 3(c) and $31.62\ meV$ in Fig. 3(d)) increase while the contributions of the other modes decrease along the distortion.
A total of 36 (and 42) vibrational modes are listed in the Table II as making significant contributions (exceeding $0.25\%$) to the JT distortion from $G_{HS}$ (or $G_{S}$) to $G_{M}$. Together, these modes account for approximately $87.0\%$ and $87.9\%$ of the total distortions, respectively, while contributing $64.7\%$ of the JT stabilization energy and $62.5\%$ of the energy barrier between adjacent minima. Importantly, most of the JT-active modes identified by the IDP analysis have also been revealed by two-dimensional electronic spectroscopy (2DES) experiments\cite{Huxter13}. The vibrational modes at $22.6\ meV$ and $189.1\ meV$, observed in the experimental vibrational bath\cite{Huxter13}, are likely associated with delocalized lattice dynamics of diamond and do not participate in the JT distortion. Vibrational modes at $106.5\ meV$ meV and $157.3\ meV$/$156.9\ meV$ closely match those observed in 2DES experiments ($105.7\ meV$ and $156.5\ meV$), as evidenced by pronounced peaks in corresponding Fourier-transform spectra (Fig. 3d of Ref.\cite{Huxter13}). The modes at $75.3\ meV$/$75.9\ meV$ and $86.5\ meV$/$86.8\ meV$ are consistent with the experimentally reported values of $75.0\ \text{meV}$ and $85.3\ \text{meV}$\cite{Carbery2024}. In addition, the modes near $64.0\ meV$ and $63.4\ meV$are in good agreement with the $65\ meV$ mode reported in earlier experiments\cite{Huxter13}. The JT-active modes obtained from the IDP method are also consistent with results from AIMD simulations\cite{Ulbricht16}. Minor deviations in the vibrational frequencies can be attributed to methodological differences: the IDP analysis is performed at zero temperature within the adiabatic approximation, whereas the AIMD simulations are carried out at finite temperatures ($10\ K$ or $77\ K$), where nonadiabatic transitions between the excited state $e_{x}$ and $e_{y}$ orbitals on different APES sheets are include\cite{Ulbricht16}.

Our calculations suggest that the observed vibrational modes in 2DES measurement are closely linked to the JT effect in the excited states of the NV center. Most of these modes actively participate in the electron-phonon coupling dynamics and play an crucial role in the JT distortion. We further find that the energy barrier between adjacent minima on the APES is sufficiently small to enable thermally activated pseudorotation. Notably, specific vibrational modes—such as the  $32.1\ meV$ mode observed in 2DES\cite{Huxter13}—appear exclusively in the distortion pathway from the saddle point to the minimum in our calculations. This result supports the existence of an allowed vibronic relaxation sequence connecting neighboring minima via saddle points (M–S–M–S–…) on the APES.  These findings are further corroborated by recent ab initio molecular dynamics (AIMD) simulations, which yield vibrational modes closely resembling those identified in the present work, thereby validating the reliability and accuracy of our computational approach. Moreover, the methodology developed here can be readily extended to investigate multimode JT effects in other defect centers, providing a general framework for studying key properties such as spin–orbital coupling, dephasing, and relaxation.

\textit{Acknowledgments} Work at Ames Laboratory was supported by the US Department of Energy, Office of Science, Basic Energy Sciences, Division of Materials Science and Engineering, including a grant of computer time at the National Energy Research Scientific Computing Centre (NERSC) in Berkeley, CA. Ames Laboratory is operated for the U.S. DOE by Iowa State University under contract \#  DE-AC02-07CH11358. J. H. Zhang was also supported by National Natural Science Foundation of China under Grant Nos.11204257, Natural Science Foundation of Hainan Province (121MS002, 122RC542), and the Research Start-up Fund Project of Hainan University (KYQD(ZR)-21066). Z. Z. Zhu was supported by the National Natural Science Foundation of China under Grant Nos. 21233004.  V. V. Dobrovitski was also supported by the research programme NWO QuTech Physics Funding (QTECH, programme 172) with project number 16QTECH02, which is (partly)  financed by the Dutch Research Council (NWO).

\bibliographystyle{unsrt} 
\bibliography{nvcentercited}

\end{document}